\documentstyle[eqsecnum,aps]{revtex}   % dos columnas
\begin{document}
\draft
\tighten
% The following line is crucial to get two-column format right
%\twocolumn[\hsize\textwidth\columnwidth\hsize\csname @twocolumnfalse\endcsname
\title{Superconductor--insulator
transition in {\sl d}-wave superconductors}
\author{A.G.Rojo$^a$ and C.A.Balseiro$^b$}
\address{$^a$The Harrison Randall Labroratory
of Physics , The University of Michigan, Ann Arbor, MI 48109-1120}
\address{$^b$Centro At\'omico Bariloche, 8400-Bariloche, Argentina}
\maketitle

\begin{abstract}
We address the question of whether an anisotropic gap $d_{x^2-y^2}$ symmetry
is compatible with localized states in the normal phase. The issue is
important in high $T_c$ superconductors where a superconductor to insulator
transition is observed in the underdoped regime, together with a number of
experiments that support $d$-wave pairing.
We find a reentrant behavior of superconductivity in the
strongly disorder phase.
\end{abstract}

\pacs{PACS numbers: 75.50.Lk, 05.30.-d. 75.40.Gb}
%\vskip2pc] \narrowtext

There is a growing body of experimental evidence in high Tc superconductors
that indicate that the pairing state is of $d_{x^2-y^{2\text{ }}}$ symmetry%
\cite{dexp}. In superconductors with an anisotropic order parameter, both
magnetic and non-magnetic impurities are pair breaking.For
$d$--wave symmetry, the effect of
non-magnetic impurities is equivalent to magnetic impurities in $s$-wave
superconductors\cite{anisot}. Perturbation theory in the impurity scattering
introduces a mean free path $\ell $ for the extended states, and the
standard treatment indicates that anisotropic superconductivity is destroyed
when $\xi /\ell \simeq 0.1$\cite{maki}, with $\xi $ the coherence length.
The situation is somewhat different for the so called extended $s$--wave
symmetry. This corresponds to an order parameter with uniform sign
 which could, in particular, vanish at certain ${\bf k}$
 directions\cite{norman}.
Point impurities are not pair--breaking in this case, but they are
``pair--weakening": for small impurity concentration $n_{\rm imp}$,
$T_c$ decreases linearly with $n_{\rm imp}$,
but the critical impurity concentration is $n_{\rm imp}^{\rm (cr)}
\rightarrow \infty$.
On
the other hand, charge dynamics in oxide superconductors is basically
two-dimensional, and it is known from the scaling theory of localization
that in two dimensions all the one particle states are localized\cite{leera}.
%, with a
%localization length $\lambda =\ell e^{k_F\ell }$\cite{leera}.
Actually, the
experimental evidence from resistivity measurements
 for low-doping
 is consistent with a divergent resistivity as
$T\rightarrow 0$ that is cut--off by the superconducting transition
\cite{takagi}.
The resistivity shows an upturn at a characteristic temperature
$T_{\rm min}$ that is apparent when $T_c < T_{\rm min}$.
Qualitatively, $T_{\rm min}$ corresponds to the temperature
scale  for which the inelastic scattering length is comparable
to the localization length. Conversly, if the {\it elastic} mean free
path is much bigger than the coherence length, then $T_c > T_{\rm min}$
the localization effects are not  important, and the
variation of $T_c $ with disorder will be given by the usual pair
breaking expressions.
In the
regime of $T_c < T_{\rm min}$,
  it is clear that  superconductivity establishes at
a temperature low enough for the effects of localization to be
of importance.
The purpose of the present work is
to present a treatment of anisotropic superconductivity that incorporates
the fact that states from which the superconducting state
is built up are
localized, and reconcile two seemingly conflicting properties:\ the observed
insulator--superconductor transition, and anisotropic pairing. We also
discuss a new prediction that emerges from our theory regarding reentrant
superconducting to insulating transition at low temperatures.

For concreteness we consider fermions on a lattice described by the
following Hamiltonian
\begin{equation}
\label{H1}H=H_0-U\sum_x\hat{\Delta }_x^{\dagger }\hat{\Delta }_x,
\end{equation}
with $H_{0\text{ }}$being a one--electron Hamiltonian that includes
disorder, with eigenstates $\varphi _\nu (x)$ and eigenvalues $\varepsilon
_\nu $. The second term in (\ref{H1}) corresponds to an instantaneous
attractive interaction with an implicit cutoff at a characteristic energy $%
\omega _D$. In order to model $d_{x^2-y^{2\text{ }}}$ symmetry we choose $%
\hat{\Delta }_x^{\dagger }$ of the form
\begin{equation}
\label{del}\hat{\Delta }_x^{\dagger }=\frac 1{\sqrt{2}}\sum_\delta \epsilon
_\delta \left( c_{x\uparrow }^{\dagger }c_{x+\delta \downarrow }^{\dagger
}-c_{x\downarrow }^{\dagger }c_{x+\delta \uparrow }^{\dagger }\right) ,
\end{equation}
with $\delta =\pm {\bf e}_1,{\bf \pm e}_2$ being the lattice versors, and $%
\epsilon _{\pm {\bf e}_1}=-\epsilon _{\pm {\bf e}_2}=1$. We argue below that
the effects of localization on the critical temperature $T_{c\text{ }}$for $%
d $--wave pairing are qualitatively the same as those for $p$--wave pairing,
an example of which is the spin polarized phase \cite{legget}, that we model
with $\hat{\Delta }_{x\sigma }^{\dagger }=\sum_\delta \epsilon _\delta
^{^{\prime }}c_{x\sigma }^{\dagger }c_{x+\delta \sigma }^{\dagger }$; $%
\epsilon _{+{\bf e}_1}=-\epsilon _{-{\bf e}_1}=1$, and $\epsilon _{+{\bf e}%
_2}=\epsilon _{-{\bf e}_2}=0$.

The critical temperature is determined by the self consistent solution of
the following linearized gap equation\cite{degennes}
\begin{equation}
\label{ker1}\Delta _x=\sum_{x^{^{\prime }}}K(x,x^{\prime })\Delta
_{x^{\prime }},
\end{equation}
where $\Delta _x=\langle \hat{\Delta }_x\rangle $, and the kernel $K,$
written in terms of the exact eigenstates of $H_{0\text{ }}$is given by
\begin{equation}
\label{ker2}K(x,x^{\prime })=U\frac T2\sum_{\omega _n}\sum_{\mu \nu \delta
\delta _{}^{\prime }}\epsilon _\delta \epsilon _{\delta ^{\prime }}\frac{%
\varphi _\mu ^{*}(x)\varphi _\nu ^{*}(x+\delta )\varphi _\nu (x^{\prime
})\varphi _\mu (x^{\prime }+\delta )}{(\varepsilon _\nu -i\omega
_n)(\varepsilon _\mu +i\omega _n)}
\end{equation}
with $T$ the temperature and $\omega _n=(2n+1)\pi T$ the Matsubara
frequencies. Also, we have taken $\hbar =k_B=1$. From now on we will take
the eigenstates as real.

We next assume that the gap is uniform, $\Delta _x=|\Delta |$, which is
justified for $\omega _{D}\gg \Delta W$, with $\Delta W$ the typical level
spacing between states within a localization length of each other. In that
case we can integrate (\ref{ker2}) over $x$ and $x^{\prime }$ and reach the
condition
\begin{equation}
\label{cond}1=\frac T2N_FU\int d\xi d\xi ^{\prime }\sum_{\omega _n}\frac{%
g(\xi -\xi ^{\prime })}{(\xi ^{\prime }+i\omega _n)(\xi -i\omega _n)}
\end{equation}
with $N_F$ the density of states at the Fermi level, and $g(\omega )$ is the
power spectrum of the operator $\hat{D}=\sum_{x,\delta }\epsilon _\delta
\left( |x+\delta \rangle \langle x|+|x\rangle \langle x+\delta |\right) ,$%
given by
\begin{equation}
\label{power}g(\omega )=\sum_\nu \overline{|\langle \mu |\hat{D}|\nu \rangle
|^2}\delta (\varepsilon _\nu -\varepsilon _F-\omega ),
\end{equation}
where the symbol represents an average over states $\mu $ such that $%
\varepsilon _\mu =\varepsilon _F$. For $s$--wave symmetry, $\hat{D}$
corresponds to the density operator $\hat{D} =\sum_{x}|x\rangle \langle x|$
, and $g(\omega )=\delta (\omega )$, since the density response is not
sensitive to scattering (in the $q=0$ limit, which is our case of interest).
This is valid even when the states $\varphi _\nu (x)$ are localized. With
this frequency dependence of $g(\omega )$ replaced in Eq. (\ref{cond}) one
obtains an equation for the critical temperature identical to that of the
pure system. This is the extension of the Anderson theorem to the case of
localized states, which was discussed by Ma and Lee using a variational
approach\cite{malee}. We conclude that under the above assumptions ($\omega
_{D}\gg \Delta W$ and uniform gap), the effects of localization on the
critical temperature are contained, through Eq.(\ref{cond}), on the
frequency dependence of spectral function of the operator that has the
symmetry of the order parameter. The function $g(\omega )$ can be calculated
diagramatically, since it is given by a two--particle bubble with bare
vertices $\gamma _{{\bf k}}^d=\cos k_x-\cos k_y$ for $d$--wave, and $\gamma
_{{\bf k}}^p=i\sin k_x$ for $p$--wave symmetry. From now on we ignore the
lattice effects, and take $\gamma _{{\bf k}}^d=\cos 2\theta _{{\bf k}}$,
which corresponds to a gap function $\Delta({\bf k})=\Delta(T)\cos 2\theta_{%
{\bf k}}$.

We first write Eq. (\ref{power}) as
\begin{equation}
\label{power2}g(\omega )=\frac 1{ 2\pi  ^2}
{\rm Re}%
\sum_{{\bf k},{\bf k}^{\prime }}\cos 2\theta _{{\bf k}}\Phi _{{\bf k},{\bf k}%
^{\prime }}(\omega )\cos 2\theta _{{\bf k}^{\prime }}\ ,
\end{equation}
with
\begin{equation}
\label{fi}\Phi _{{\bf k},{\bf k}^{\prime }}=\langle G^{{\rm R}}({\bf k},{\bf %
k}^{\prime };\varepsilon _F)G^{{\rm A}}({\bf k}^{\prime },{\bf k,\varepsilon
}_F^{}+\omega )\rangle \ ,
\end{equation}
where now $\langle \ldots \rangle $ denotes impurity average. We follow the
work of Vollhardt and W\"olfle (VW)\cite{wolfe}, and compute $g(\omega )$
within the self--consistent theory of localization. We prove that the
frequency dependence of (\ref{power2}) is essentially the same as that of
the conductivity. The quantity $\Phi _{{\bf k},{\bf k}^{\prime }}$ obeys the
Bethe--Salpeter equation
\begin{equation}
\label{bethe}\Phi _{{\bf k},{\bf k}^{\prime }}(\omega )=G_{{\bf k}}^{{\rm R}%
}G_{{\bf k}}^{{\rm A}}\delta _{{\bf k},{\bf k}^{\prime }}+G_{{\bf k}}^{{\rm R%
}}G_{{\bf k}}^{{\rm A}}\sum_{{\bf k}^{\prime \prime }}U_{{\bf k},{\bf k}%
^{\prime \prime }}(\omega )\Phi _{{\bf k}^{\prime \prime },{\bf k}^{\prime
}}(\omega )\ ,
\end{equation}
with $U_{{\bf k},{\bf k}^{\prime \prime }}(\omega )$ the {\it irreducible }%
vertex function. In Anderson localization, {\it single}-particle quantities
(e.g. the density of states) are smoothly varying functions of disorder. It
is then reasonable to approximate the self energy $\Sigma $ by the lowest
order result in the impurity scattering $U_0,$ and use the Green's functions
in the form
\begin{equation}
\label{green}G_{{\bf k}}^{{\rm R}}(\varepsilon )=\frac 1{\varepsilon
-\varepsilon _{{\bf k}}+\frac i{2\tau }}\ ,
\end{equation}
with, $1/\tau =2\pi N_FU_0,$ and $G_{}^{{\rm R}}=(G^{{\rm A}})^{*}.$ We have
assumed a $\delta $ correlated disordered potential $u(x)$, such that $%
\langle u(0)u(x)\rangle =u_0^2n_{{\rm imp}}\delta (x)=U_0\delta (x)$. Using
this expression for the Green's functions we can write the Bethe Salpeter
equation as a kinetic equation in the form
\begin{equation}
\label{kin}(\omega -{\frac i\tau })\Phi _{{\bf k},{\bf k}^{\prime }}=-\Delta
G_{{\bf k}}\left[ \delta _{{\bf k},{\bf k}^{\prime }}+\sum_{{\bf k}^{\prime
\prime }}U_{{\bf k},{\bf k}^{\prime \prime }}\Phi _{{\bf k}^{\prime \prime },%
{\bf k}^{\prime }}\right] \ ,
\end{equation}
with $\Delta G_{{\bf k}}\equiv G_{{\bf k}}^{{\rm R}}-G_{{\bf k}}^{{\rm A}}$.
If we replace in (\ref{kin}) the irreducible vertex by the bare vertex $U_0$%
, we obtain
\begin{equation}
\label{gmet}g(\omega )=\frac{N_F}{4\pi }\frac \tau {1+(\omega \tau )^2}\ .
\end{equation}

A very simmilar expression was obtained before in a treatment of the Raman
response in the $l=2$ channel in the presence of impurities\cite{zawa}.
However, since the Raman response $R(\omega )$ is given by a correlation
function, there is an additional factor of $\omega $ and $R(\omega )=\omega
g(\omega )$. Inserting (\ref{gmet}) in (\ref{cond}), we obtain the well
known expression for the critical temperature variation\cite{radtke} $-\ln
(T_c/T_{c0})=\Psi (1/2+1/4\pi \tau T_c)-\psi (1/2)$.

%The signature of localization in this approach is the divergence
% of the irreducible vertex at low frequencies.
Following VW, we observe that since $\Delta G_{{\bf k}}$ is srongly peaked
at $k=k_F$, the dependence of $\Phi_{{\bf k}, {\bf k}^{\prime}}$ on the
magnitude of the wave vectors will be dominated by $\Delta G_{{\bf k}}$. We
define $\Phi_{{\bf k}}= \sum _{{\bf k }^{\prime}} \Phi_{{\bf k},{\bf k}%
^{\prime}}\cos 2 \theta_{{\bf k}^{\prime}} $, and extract the angular
dependence in {\bf k} using a Legendre expansion in which we keep up to the $%
l=2$ term:
\begin{equation}
\label{fii}\Phi_{{\bf k}}={\frac{{\Delta G_{{\bf k}}}}{{-2\pi i N_F}}}\sum_{%
{\bf k}^{\prime}}\left[ 1+2\cos \theta_{{\bf k}} cos \theta_{{\bf k}%
^{\prime}} +2\cos 2\theta_{{\bf k}} cos 2\theta_{{\bf k}^{\prime}}\right]%
\Phi_{{\bf k}^{\prime}}\ .
\end{equation}

Multiplying (\ref{kin}) by $\cos 2\theta _{{\bf k}}\cos 2\theta _{{\bf k}%
^{\prime }}$, summing over ${\bf k},{\bf k}^{\prime }$, and using (\ref{fii}%
), we obtain
\begin{equation}
\label{res1}\sum_{{\bf k},{\bf k}^{\prime }}\cos 2\theta _{{\bf k}}\Phi _{%
{\bf k},{\bf k}^{\prime }}\cos 2\theta _{{\bf k}^{\prime }}={\frac{-{i\pi N_F%
}}{{\omega -M(\omega )}}}\ ,
\end{equation}
where $M(\omega )$ is a ``$l=2$ relaxation kernel'', given by
\begin{equation}
\label{kernel}M(\omega )={\frac{{i}}{{\tau }}}+{\frac{{i}}{{\pi N_F}}}\sum_{%
{\bf k},{\bf k}^{\prime }}\cos 2\theta _{{\bf k}}\Delta G_{{\bf k}}U_{{\bf k}%
,{\bf k}^{\prime }}\Delta G_{{\bf k}^{\prime }}\cos 2\theta _{{\bf k}%
^{\prime }}\ .
\end{equation}

The structure of $M(\omega )$ is very simmilar to that of the current
relaxation kernel. Note that in deriving this expression we have neglected
terms that mix different angular dependences in $M(\omega )$, and that give
rise to factors $\cos m\theta _{{\bf k}}\cos 2\theta _{{\bf k}^{\prime }}$,
with $m=0,1,2$. These terms do not appear in the treatment of VW. Neglecting
these terms is justified, since we are anticipating the inclusion of the
contribution to the irreducible vertex that originate the divergence of $%
M(\omega )$. The infrared divergence of $M$ comes from the maximally crossed
(MC) diagrams\cite{langer}, which are irreducible, and contribute with
\begin{equation}
\label{MC}U_{{\bf k},{\bf k}^{\prime }}^{{\rm MC}}={\frac{{iU_0/\tau }}{{%
\omega +iD_0({\bf k}+{\bf k}^{\prime })^2}}}\ ,
\end{equation}
for ${\bf k\simeq }-{\bf k}^{\prime }$, and with $D_0$ the bare difussion
constant. Due to this divergence we can take ${\bf k}={\bf k}^{\prime }$ for
the angular integral and the ``off--diagonal'' contributions vanish due to
orthogonality. Using the above $U_{{\bf k},{\bf k}^{\prime }}^{{\rm MC}}$ in
(\ref{kernel}) we obtain the logaritmic low--frequency divergence $M(\omega
)\sim \log \omega $ that is familiar from perturbation theory of the
conductivity\cite{klemi}. Since the low frequency $l=2$ kernel is
essentially the same as the current relaxation kernel (the angular
integrations give the same result), it will still be related to the
difussion constant $D(\omega )=iD_0[M(\omega )\tau ]^{-1}$. This allows us
to go beyond perturbation theory and determine $M(\omega )$ self
consistently through the equation
\begin{equation}
\label{self}M(\omega )={\frac{{i}}{{\tau }}}-{\frac{{2}}{{\tau }}}\sum_{{\bf %
k}}{\frac{{1}}{{\omega -k^2D_0[M(\omega )\tau ]^{-1}}}}\ .
\end{equation}

Equation (\ref{self}) can be solved for low frequencies, giving
\begin{equation}
\label{mfin}M(\omega )={\frac{{i}}{{\tau }}}-{\frac{{\omega _0^2}}{{\omega }}%
}
\end{equation}
and
\begin{equation}
\label{gfin}g(\omega )=\frac{N_F}{4\pi }\frac \tau {1+(\widetilde{\omega }%
\tau )^2}\ ,\
\end{equation}
with $\widetilde{\omega }=\omega +\omega _0^2/\omega $.
 The  characteristic frequency
$\omega_0$ is finite
in the localized phase, and is given by
 $\omega_0=v_F/(\sqrt{2}\lambda)$, with $lambda$ the localization length. This
result implies
an  expression for the dependence of the critical temperature, which
can be considered as a generalization of the Abrikosov--Gorkov--Maki (AGM)
formula for localized anisotropic superconductors. Our result is
then
\begin{equation}
\ln({{T_{c0}}\over T_c})=
-\ln({T_{c0}\over \omega_D}) {2 \tau \over \tau^-}
+
{ \tau \over \tau^+}
\left[
\psi({{1}\over{2}}+\rho_+)-
\psi({{1}\over{2}})\right]
%\end{equation}
%\begin{equation}
-
{ \tau \over \tau^-}
\left[
\psi({{1}\over{2}}+\rho_-)-
\psi({{1}\over{2}})\right] ,
\end{equation}
where ${ 1/ \tau^{\pm}}=\sqrt{\omega_0^2+(1/2\tau)^2}
\pm 1/\tau$, and $\rho_{\pm}=1/4\pi\tau^{\pm}T_c$.

The above expression gives the change in critical temperature as a function
of $\tau $ and $\omega _0$.
In contrast
with the AGD formula,
in our case
the relative change in $T_c$ is
dependent of the cutoff frequency $\omega_D$.
In order to visualize the effects of localization on $T_c$ predicted
by our treatment, in Figure 1 we show plots of $T_c$  as a function of
$\tau $, treating $\omega _0$ as an independent parameter. In principle
these two quantities are not independent, but the plot emphasizes the
fact that when $T_c/T_{c0} < 0.1 $ there is a reentrant behaviour to a normal
phase at very low temperatures. We note that the reentrant behaviour
correponds to a parameter range that is realistic in high Tc oxides.
The curves of Figure 1 show that there is reentrant behaviour when
$\omega_0 \tau \sim 1$. From the self consistent theory,
$\omega_0 \tau=\ell/{\sqrt{2} \lambda}$, where $\lambda $ is the localization
length. On the other hand $ {\lambda \over \ell} = (e^{2k_F \ell} -1)^{1/2}$.
In high $T_c$ oxides, since the particle density is small,
$k_F \ell$ can be of order one, and $\omega_0 \tau \sim 1$.
We conclude that localization effects are important in high Tc
oxides, and  that $d$--wave superconductivity can coexist
with Anderson localization.

Finally, we discuss briefly the mechanism underlying the reentrant
behavior in $T_c$. The reentrance to a normal state as temperature
decreases can occur if the entropy of the superconducting phase
is higher  than the entropy of the normal state at very low
temperatures. This can happen if the localization length is
of the order of the mean free path. In this regime one can estimate
the density of states $\rho(E)$ of the quasiparticles by computing the
quasiparticle energies $E_{\nu}$ as a correction of the {\it particle}
energies $\varepsilon_\nu $
 in second order in the pairing
interaction\cite{degennes}.
%\begin{equation}
%E_{\nu}= \varepsilon_\nu + \sum_\mu {M_{\mu \nu}^2\over
% \varepsilon_\nu  + \varepsilon_\mu},
%\end{equation}
%with
%\begin{equation}
%M_{\mu \nu}^2=2(U\Delta)^2\sum_{x\delta}\sum_{x'\delta'}
%\epsilon_{\delta} \epsilon_{\delta'}
%\varphi _\nu (x)
%\varphi _\mu (x+\delta)
%%\varphi _\nu (x')
%\varphi _\mu (x'+\delta ').
%\end{equation}
%
Since  states that are degenerate have negligible overlap in
the localized phase, we can extract the typical correction to
the quasiparticle energy from $g(\omega)$, and obtain
\begin{equation}
\rho(E) ={N_0 \over \left| {\partial E\over \partial \varepsilon}\right|}=
{N_0 \over 1+{\partial\over \partial \varepsilon}
\left. \int d\omega {g(\omega) \over  \varepsilon +\omega}\right|_{\varepsilon
=E}}.
\end{equation}
The result is then that the density of states at
the fermi energy can be bigger than the normal state density of
 states, and the superconducting entropy is higher at low
 temperatures. The behavior of $\rho(E)$ is nonmonotonic in $E$ and
 on e gets two transitions as temperature increases.

In comparing with experiments
one should be able to isolate the effects of hole doping
and the effects of disorder.
 Recent experiments on  Bi$_2$Sr$_2$Ca(Cu,Co)O$_{8+y}$\cite{quitmann}
it is shown that $T_c$ shows a drastic decrease with
increasing disorder and the resistivity shows a minimum
suggesting Anderson localization.
{}From angle resolved photoemission it is concluded that the carrier
concentration is unchanged, as expected since Co is a $2^+$ ion. Hall
effect measurements are also indicate that the carrier concentration
does not change\cite{maeda}.
{}From Ma and Lee's theory\cite{malee}
one does
not expect a big decrease in $T_c$ if the symmetry is $s$-wave,
since Anderson theorem still holds if the mobility edge
crosses the chemical potential. The fact that $T_c$
is  changing in the ``coexistence region" between the insulating and
superconductivity phases, can be better understood if
the symmetry is $d$--wave.
The reentrant behaviour has not been measured to our knowledge, and
it constitutes a well defined experimental test of the present
theory.

 In summary, we have shown that $d$--wave superconductivity
 is compatible with Anderson localization of the one--particle
 states.
 In addition, out treatment extends previous calculation of the
 Raman response in disordered systems to the
 localized phase.
 We also made the interesting prediction of
 reentrance to the normal phase as temperature is lowered
 in the superconducting phase, when disorder is strong.

\acknowledgments
A.G.R would like to thank C. Balseiro and G. Dussel for their kind
hospitality during stays at the Centro At\'omico Bariloche and Universidad
the Buenos Aires, where part of this work was done, and to L. Jaimovich from
the
Universidad de Tucum\'an, Argentina. We also thank K.Levin, Eduardo Fradkin and
J. Rasul for discussions.

\begin{figure}
\caption{
Critical temperature vs. inverse scattering time $\Gamma=1/4\pi \tau  T_{c0}$
for fixed values of $\omega_0=0.$ (solid), $0.1$ (dashed), $0.2$ (long dashed)
$0.25$ (dot--dashed).}
%in the c-direction (b). The three curves correspond to increasing values of
%$V_2/t_{bot}$ (amplitude of dynamic off diagonal disorder): 0 (continuos), 1.
%%(dashed)
%and 2. (short dashed). The inset corresponds to $t_{bot}=0$ and different
%values of $V_2 /t_{\|}$.}
\end{figure}

\end{document}